\documentclass[runningheads]{llncs}
\usepackage[T1]{fontenc}
\usepackage{graphicx}
\begin{document}
\title{AISSA Demo: AI-based Student Slides Analysis Tool for Automated Grading and Feedback}
\titlerunning{AISSA Demo: AI-based Student Slides Analysis}
%
\author{Alvaro Becerra\inst{1}\orcidID{0009-0003-7793-2682} \and
Diego Gomez\inst{1} \and \\
Ruth Cobos\inst{1}\orcidID{0000-0002-3411-3009} \and Julian Fierrez\inst{2}\orcidID{0000-0002-6343-5656}}
\authorrunning{A. Becerra et al.}
\institute{GHIA, School of Engineering, Universidad Autónoma de Madrid, Spain \and BiometricsAI, School of Engineering, Universidad Autónoma de Madrid, Spain}
%

%
\maketitle              
\begin{abstract}
We present an AI-based web-oriented tool designed to support formative feedback for oral presentation slides in higher education: AISSA. It allows students to upload their slides before presentation and automatically receive rubric-based quantitative scores and qualitative feedback generated by large language models (LLMs). The tool analyses slide-level features and content using a teacher-defined rubric and delivers feedback through interactive Learning Analytics dashboards. These dashboards enable students to visualize performance indicators, inspect feedback in context, and reflect on strengths and areas for improvement, while teachers can review automated assessments, provide their own evaluations, and monitor student engagement with feedback.

\end{abstract}
\section{Introduction}

Feedback plays a central role in learning by helping students identify the gap between their current performance and learning goals, thus supporting reflection, self-regulation, and improvement \cite{hattie2007power,schartel2012giving}. However, providing effective feedback in higher education remains challenging, as comments are often vague, insufficiently specific, poorly aligned with future tasks, or delivered too late to be useful \cite{henderson2019challenges}. These difficulties are further intensified by increasing teaching workloads and the need to support large and diverse cohorts of students.

To address these limitations, recent research has explored the use of artificial intelligence (AI) and large language models (LLMs) to support scalable assessment and feedback practices. Automated feedback systems can provide timely and consistent support, while advances in AI and natural language processing have expanded their potential beyond numerical scoring to formative feedback that promotes reflection and self-improvement \cite{cavalcanti2021automatic,gao2024automatic}. At the same time, the use of generative AI in education raises important issues related to transparency and responsible adoption \cite{beardsley2025learning}. Recent studies have highlighted the potential of LLMs for automated evaluation in contexts such as multimodal evaluation, open responses, and rubric-based essay scoring \cite{shin2025can,srivastava2025learnlens,mello2025empowering}, as well as for generating assessment activities and evaluating learner responses with human oversight \cite{askarbekuly2024llm}. In parallel, Learning Analytics dashboards (LADs) \cite{becerra2025m2lads} help make assessment output more interpretable and actionable for teachers and students \cite{becerra2025review,verbert2013learning,topali2025designing}. Even before the rise of LLMs, multimodal learning analytics systems had already shown the feasibility of automatically evaluating presentation slides using predefined visual and textual features \cite{ochoa2024openopaf}.

This paper presents a web-based system that combines LLMs and LADs to support scalable rubric-based assessment, and formative feedback for oral presentations through the automatic analysis of presentation slides: AISSA.\footnote{Demonstration of the full AISSA workflow at: \url{https://youtu.be/QYoZTVH-\_YU}}

\section{AISSA: AI-based Student Slides Analysis}
AISSA was developed following a human-centered design approach \cite{topali2025designing}, which involved iterative consultations with teachers and students to align it with educational needs and assessment practices. AISSA allows students to upload their slides before an oral presentation and receive quantitative scores together with qualitative actionable feedback generated by LLMs and delivered through LADs.

AISSA is composed of three modules: the Feedback Generation Module (see Sect.~\ref{ss:feedback_module}); the Management Module, which connects the tool to its data storage infrastructure, including SQL and MongoDB databases and the file system; and the Visualization Module (see Sect.~\ref{ss:visualization_module}).

\subsection{Feedback Generation Module}\label{ss:feedback_module}
The feedback generation module processes slides uploaded by students by extracting slide-level features such as text length, font type, number of lines, images, and other layout characteristics. These features, together with the slide content, are used to build a prompt that is sent to ChatGPT 5.2 \cite{becerra2024generative}.

The system uses a teacher-defined rubric composed of evaluation criteria assessed on a five-point Likert scale with explicit descriptors. The default rubric, designed by oral presentation experts at UAM, includes nine criteria related to slide quality, such as text readability, contrast, and visual balance. Teachers can adapt the rubric criteria through the dashboard, and these modifications are automatically incorporated into the LLM prompt used for feedback generation.

The model returns a numerical score and a brief justification for each item of the rubric, as well as qualitative feedback structured into three parts: strengths, areas for improvement, and suggestions for improvement \cite{hattie2007power,becerra2025enhancing}. AISSA then validates the generated output by checking constraints such as length and excluded terms and automatically regenerates the feedback when necessary. Once validated, the feedback is stored and displayed to students through the dashboards.

To support ethical and transparent use, students are informed in advance that their slides will be uploaded to an external GenAI server for automated analysis. They are also instructed not to include personal or sensitive information in the submitted materials. The platform and feedback are available in English and Spanish, regardless of the presentation language.

\subsection{Visualization Module}\label{ss:visualization_module}

\begin{figure}[t]
    \centering
    \includegraphics[width=0.75\linewidth]{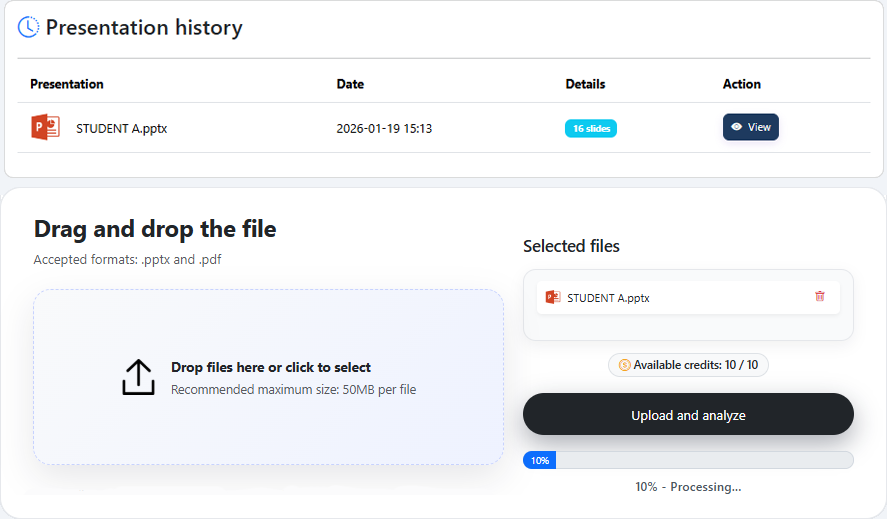}
    \includegraphics[width=0.75\linewidth]{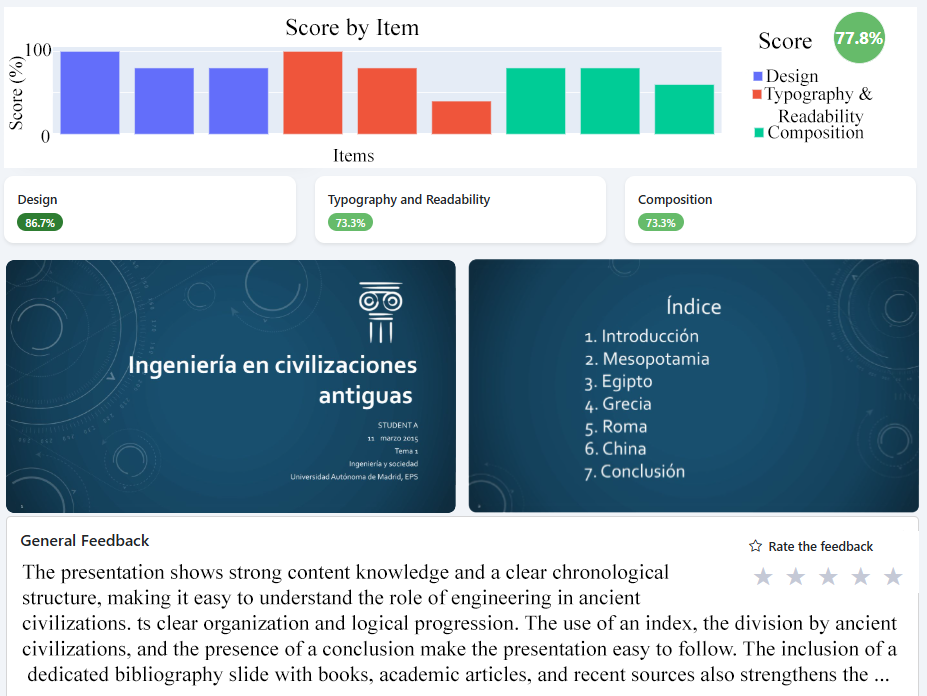}
    \caption{AISSA \textbf{Student Dashboard}: Slides upload (top) and Feedback (bottom).}
    \label{fig:upload}
\end{figure}

Several Learning Analytics dashboards were designed and implemented using the Plotly Dash framework. 
Each student is provided with an individual \textbf{Student Dashboard}. Within this dashboard, students can upload their presentation slides for analysis and feedback generation prior to an oral presentation and review previously submitted slides through a historical view (see Fig.~\ref{fig:upload} top).

Once the slides are uploaded and processed (see Sect.~\ref{ss:feedback_module}), the generated feedback is presented to the students through the dashboards. Students can view and navigate their uploaded slides alongside feedback, allowing them to contextualize scores and comments with respect to the specific elements of their presentation (see Fig.~\ref{fig:upload} bottom). The dashboard also provides access to the evaluation rubric used by the tool, along with a set of teacher-authored tips for creating effective oral presentation slides. Finally, graphical visualizations of the scores obtained for each rubric item help students easily identify strengths and areas for improvement by different presentation criteria.

To support reflection and continuous improvement of the tool, students are also given the opportunity to evaluate the perceived quality and coherence of the generated feedback. This feedback is collected and used to inform future refinements of the tool and the underlying AI-based feedback generation process.

\begin{figure}[t]
    \centering
    \includegraphics[width=\linewidth]{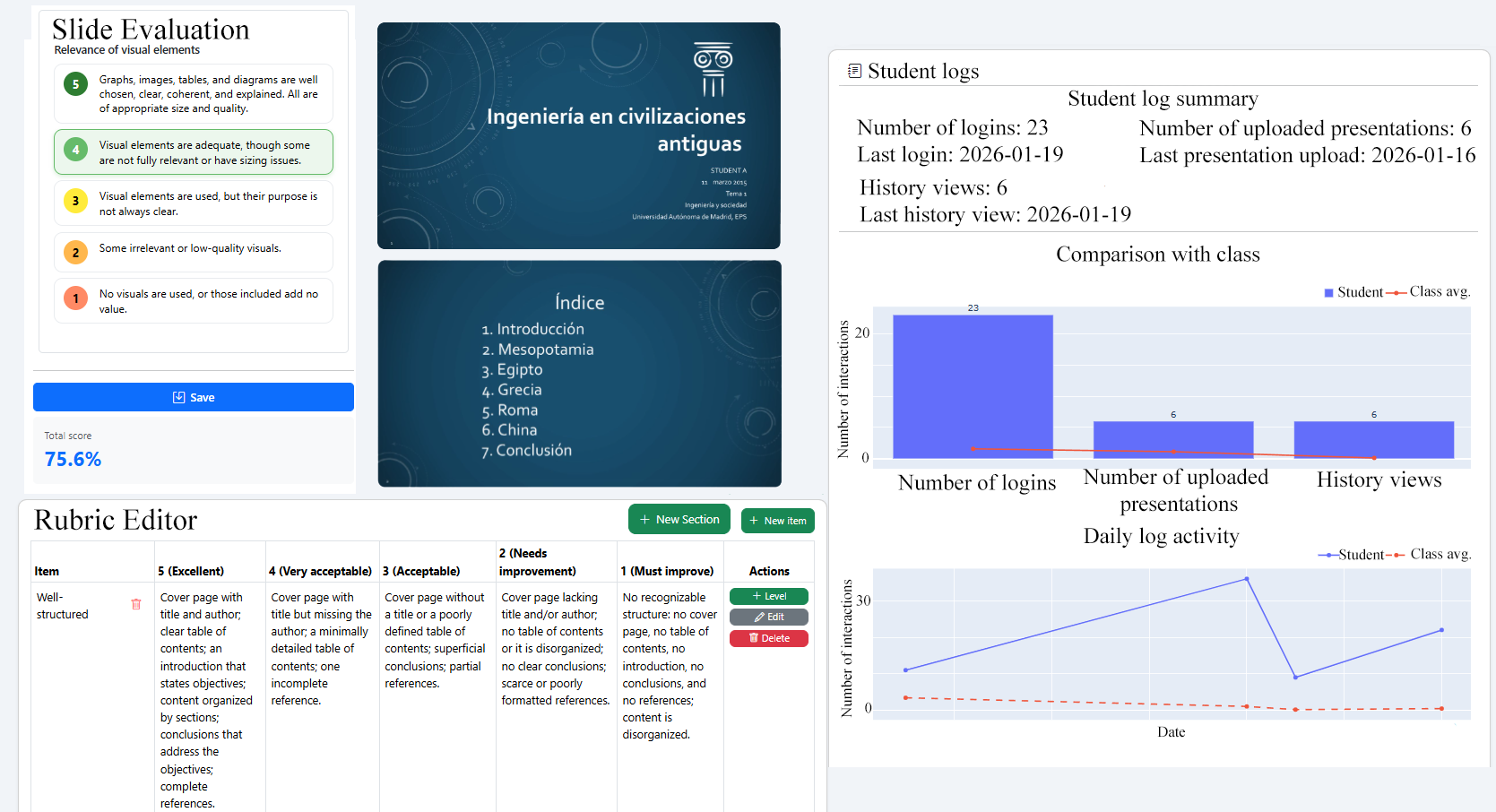}
    \caption{AISSA \textbf{Teacher Dashboard}.}
    \label{fig:teacher}
\end{figure}

The \textbf{Teacher Dashboard} (see Fig.~\ref{fig:teacher}) allows teachers to access and review the automatic evaluations generated by the system for the presentation slides uploaded by the students, including direct access to each student’s individual dashboard. In addition to inspecting AI-generated feedback, teachers can perform their own slide assessment by completing the same evaluation rubric used by the tool, which they can also edit and customize to adapt the evaluation criteria to different presentation contexts and learning objectives. When a teacher-provided evaluation is available, students are also shown the corresponding quantitative scores alongside the system-generated feedback, enabling a direct comparison between automated and human assessment.

Beyond assessment functionalities, the teacher dashboard provides access to interaction logs that capture the engagement of students with the tool. These logs are aggregated and presented through summary indicators, such as the number of times a student has accessed the dashboard, reviewed the generated feedback, or interacted with their uploaded slides. This information offers teachers insight into the patterns of participation of students. They can also edit the tool rubric. The tool also includes an administrator dashboard. Through this dashboard, teachers can create, modify, and remove students, teachers, and courses.

\section{Conclusion and Future Work}
This paper presented AISSA (``AI-based Student Slides Analysis'' tool), a web-based tool that combines LLMs and Learning Analytics dashboards to provide scalable, rubric-based feedback on oral presentation slides. By integrating automated scoring, structured qualitative feedback, and interactive visualizations for both students and teachers, AISSA addresses key challenges related to feedback quality, timeliness, and scalability in higher education.

AISSA is currently being piloted at our institution with approximately 150 undergraduate and master's students within the MOSAIC-F framework \cite{becerra2025mosaicf}, which aims to enhance oral presentation skills through personalized and data-driven feedback. The preliminary results of this pilot usability, based on the System Usability Scale (SUS) administered to 30 students, yielded an average score of $83.38$, indicating excellent perceived usability.

As future work, we plan to conduct a systematic evaluation of AISSA using presentation slides from the SOPHIAS dataset \cite{becerra2026sophias}, comparing the automated scores generated by the system with the evaluations provided by expert professors from our institution. This analysis will examine the extent to which AISSA aligns with human assessment patterns, both in terms of statistical differences and absolute agreement. In particular, our goal is to explore similarities and discrepancies at the criterion-level between the dimensions of the rubric to better understand the strengths and limitations of automated assessment.

Future research will also: 1) explore the perceived usefulness, trust, and pedagogical value of the generated feedback; 2) compare different AI models, including commercial LLMs and local or open-source alternatives, to determine which approaches align the most closely with human evaluations and are the most suitable for educational settings; 3) investigate the use of multimodal data to improve feedback.

\subsubsection{Acknowledgements.}

Support by Cátedra ENIA UAM-VERIDAS en IA Responsable (NextGenerationEU PRTR TSI-100927-2023-2), M2RAI (PID2024-160053OB-I00, MICIU/FEDER), Emotional (UAM AVANZA 2026) and SNOLA (RED2022-134284-T). A. Becerra is funded by an FPI from the Comunidad de Madrid (PIPF-2024/COM-34288).
    
\bibliographystyle{splncs04}
\bibliography{sample-base}
\end{document}